# Cities and political violence in West Africa

8/4/2026

Steven M. Radil, University of Florida, steven.radil@ufl.edu
Olivier J. Walther, University of Florida, owalther@ufl.edu

**Abstract**

This paper examines how political violence in West Africa is distributed between urban agglomerations and their surrounding areas using spatially disaggregated conflict and population data. Covering seventeen countries from 2012 to mid-2025, the analysis shows that political violence remains clearly tied to cities and their peripheries with the majority of violence occurring within and around urban agglomerations. Large cities record the highest number of incidents within their boundaries, while smaller towns, which are more frequently situated near borders, typically experience higher levels of violence outside their boundaries. However, urban-centric violence has also declined since 2012, indicating a progressive ruralization of conflict. These findings provide important regional context to the "urbanization of conflict" narrative by showing that West African insurgent movements use both urban and rural environments strategically. They also underscore the ongoing vulnerability of peripheral populations and the need for mobile and community-based security approaches outside major cities. Finally, they highlight the importance of strengthening border infrastructure, transport networks, and civic institutions as tools of both development and security.

**Acknowledgements:** The authors thank Inhoi Heo for his help with the Africapolis data.

## Introduction: changing violence scheme in urban territories

West Africa has seen a resurgence of armed conflict since the late 2000s, mainly due to the emergence of violent extremist organisations and separatist groups. Conflicts between these armed groups and government forces have produced a very specific geography of political violence in the region. On the one hand, certain places, such as small towns located on the borders of Burkina Faso, Mali, and Niger, tend to be much more affected by violence than others. On the other hand, the type of localities most affected by political insecurity have varied greatly over time. The conflicts in the Central Sahel and Lake Chad basin, which mainly targeted urban centres in the early 2010s, have spread to rural areas over the past decade. In recent years, certain Sahelian towns and cities, such as Djibo (Burkina Faso), Kayes and Bamako (Mali) have been subject to economic blockades by Jihadist groups, suggesting a new, more urban phase of conflict.

Previous studies have shown that urbanisation did not necessarily lead to a larger share of violence in urban areas in Sub-Saharan Africa, challenging the "urbanisation of conflict" hypothesis (see, among others, Elfversson et al., 2023). In West Africa, especially, most battles between armed groups, remote attacks using drones, air strikes or improvised explosive devices, and violence against civilians tend to affect rural areas, lower population density areas surrounding medium-sized cities, and smaller poorly accessible cities in border regions (OECD/SWAC, 2023).

However, previous research has focused on the region's population density and distribution as a proxy for cities and urban agglomerations. The paper fills this gap by specifically examining how political violence is distributed within urban agglomerations and beyond. Combining disaggregated data on political violence from the Armed Conflict Location & Event Data (ACLED) project and urban population data from OECD's Africapolis, the paper discusses three crucial questions for the region's future trajectory.

First, the analysis considers how much violence is happening within the geographic footprint of urban agglomerations relative to their periphery. In other words, is urban violence mainly located within the built-up core, as defined by Africapolis, or in the suburbs? To address this question, the paper develops a ratio measurement of urban security that compares the violence surrounding a city with that occurring within urban centres. Next, the paper examines what proportion of violence is associated with types of cities within the urban hierarchy. Finally, the paper examines how these patterns of violence observed in West African vary by country, over time, and within border regions.

Political violence in West Africa remains overwhelmingly tied to cities and their peripheries. About 85 percent of all violent events occur within 30 km of an urban agglomeration, confirming that cities and their peri-urban peripheries are central to the geography of insecurity. Roughly 40 percent of events occur inside city limits, with another 44 percent in adjacent peri-urban zones (defined as areas within 30 km of a city). Yet this urban concentration has changed over time. Since 2012, the share of violence taking place inside cities has fallen sharply, indicating a progressive ruralization of conflict. Violence that once centred in large cities like Maiduguri or Bamako has increasingly dispersed into hinterlands and border regions, where states exercise weaker territorial control.

Despite this diffusion, large cities remain the primary sites of organized political violence. They record the highest number of incidents within their boundaries, reflecting their demographic weight and strategic importance as political, economic, and military centres. In contrast, smaller towns, especially those near borders, experience more violence in their surrounding areas than within their cores. Border proximity significantly influences this pattern. Cities located within 40 km of an international boundary are much more likely to face external rather than internal violence, as insurgent activity and state counter-operations spill across frontier zones.

The findings have three major implications. First, the shifting geography of violence challenges the traditional "urbanization of conflict" narrative by showing that insurgent movements use both urban and rural environments strategically. Second, the increasing ruralization of warfare underscores the vulnerability of peripheral populations and the need for mobile and community-based security approaches outside major cities. Third, the persistence of violence around border towns highlights the importance of strengthening border infrastructure, transport networks, and civic institutions as tools of both development and security. Addressing these challenges requires integrated urban and regional planning that links cities more effectively to their rural and border environments.

## More urban violence?

The literature on the relationship between cities and conflict addresses two major questions for the geography of political violence: are cities more likely to experience conflict than rural areas, and to what extent does violence spread between them over time?

### *Violence in cities and rural areas*

Cities and rural areas are both vulnerable to political violence, but for different reasons. In cities, armed conflict often aims to seize control of state symbols and institutions such as parliaments or presidential offices (Beall et al., 2013). Urban areas are also prime military targets because they host government and army headquarters. The concentration of sovereign functions explains why capturing the capital can determine conflict outcomes (Goodfellow and Jackman, 2023), as seen during the Second Ivorian Civil War in 2011. Where national populations are concentrated in a few large cities, states often have only marginal control over their peripheries, fuelling rural grievances (Nedal et al., 2020).

Urban density also facilitates mobilization against the state. Large populations can enable the organization of opposition groups, while higher education levels can lead to the creation of political, civil, and religious movements that seek reform. Consequently, large African cities experience more protest and electoral violence than smaller towns (Dorward and Fox, 2022; Elfversson, 2025). Urban cosmopolitanism also attracts displaced (Büscher, 2020) and marginalized groups (Golooba-Mutebi and Sjögren, 2017) who may compete for limited urban resources (Østby, 2016).

However, evidence linking population growth to political unrest is limited. Studies find little correlation between urbanization and conflict (Buhaug and Urdal, 2013). Globally, the share of armed conflict in cities has declined since 1989 (Elfversson and Höglund, 2021). In sub-Saharan Africa, urban growth correlates with unrest mainly in peri-urban zones (Gizelis et al., 2021; Fox and Bell, 2016; Kniknie and Büscher, 2023; Elfversson et al., 2023; Radil et al., 2023). Only in North Africa has conflict and protest become increasingly urban (Dorward, 2024). These results challenge the "urbanization of conflict" hypothesis that political violence is shifting toward cities (Kaldor and Sassen, 2020).

The weak link between urbanization and violence partly reflects the appeal of rural areas to armed groups (Hendrix, 2011). Rural regions provide vital resources, such as minerals, livestock, crops, and recruits, and are often less controlled by states due to low population density and distance from power centres. Insurgencies thus thrive in remote borderlands, which offer logistical havens (OECD/SWAC, 2022; Radil et al., 2022).

The more difficult question concerns how violence moves between cities and rural areas. Maoist theory holds that rebellions emerge in rural regions before seizing cities. While influential in anti-colonial struggles such as Algeria's (Peterson, 2024), this model reflects the specific Chinese context of peasant-landowner conflict. In Africa, rebel groups often lack agrarian or political reforms to offer rural populations (Mkandawire, 2002). A framework tailored to African conditions is thus required to explain how violence circulates between urban and rural settings.

*Cities and conflicts in West Africa*

The geography of conflict in West Africa suggests that armed movements alternate between exploiting urban and rural resources. The Malian conflict began with attacks on towns before spreading into the countryside (Retaillé and Walther, 2013; OECD/SWAC, 2020), and recent years have seen growing attacks on roads and urban blockades. In the Lake Chad basin, Boko Haram has been documented as following a similar pattern for years, oscillating back and forth between Maiduguri and surrounding rural zones (Thurston, 2018).

Two factors drive these oscillations. First, urban violence depends on states' ability to project power beyond major cities. In West Africa, where effective state control is limited, rural insecurity can therefore grow unchecked. Second, armed groups pursue their own spatial strategies regarding cities. Jihadist movements, for instance, tend to view cities both as morally corrupt spaces that need their control and as strategic centres for imposing religious order (Thurston, 2020).

Understanding the role of urban and rural areas in the regions' conflicts therefore requires a broader temporal perspective. The "spatial conflict life cycle" (Walther et al., 2025) conceptualizes conflicts as regional events that emerge, spread geographically, and ultimately resolve over time. West Africa has experienced several such cycles since the Cold War: one in the 1990s around the Gulf of Guinea, and another since the 2010s beginning in the Sahel (OECD/SWAC, 2025). In each cycle, rural violence increased as armed groups extended their reach, explaining the ongoing ruralization of violence across the region (OECD/SWAC, 2023).

Given the weakness of state militaries and the mobility of insurgent actors, this ruralization could eventually reverse. Historical analogies from Somalia (1991) and Syria (2024) suggest that future jihadist offensives against major West African cities cannot be excluded. This paper contributes to documenting that evolution by examining how violence has been distributed between urban agglomerations and their peripheries since the mid-2010s.

The paper extends prior research on West African violence in three ways. First, it adopts an explicitly spatial approach that differentiates city types by size, rather than relying on categorical urban–rural distinctions common in quantitative studies, such as those conducted by OECD/SWAC (2023) in previous studies. Second, it disaggregates results to assess variation by country, year, and peripheral border regions. Third, the paper focuses on politically motivated violence rather than unrest and riots, which are already well documented by Dorward (2024).

**Data and methods: combining conflicts, population and urbanization spatial data**

To assess these questions, two different types of data were combined and analysed across 17 countries between 2012 and mid-2025: Benin, Burkina Faso, Cameroon, Chad, Côte d'Ivoire, Gambia, Ghana, Guinea, Guinea-Bissau, Liberia, Mali, Mauritania, Niger, Nigeria, Senegal, Sierra Leone, and Togo. First, conflict data from ACLED was used to document where violence has happened in the region. Second, OECD/SWAC's Africapolis dataset, which estimates the population size and spatial extent of populated places in the region, was used to identify which urban agglomerations have been impacted by violence.

*ACLED*

ACLED provides a comprehensive, disaggregated, and georeferenced record of political violence and protest events across Africa since 1997. Each event in the database is coded with detailed spatial and temporal information, including the date, location, actors involved, and the type of violence.

The dataset distinguishes among eight categories of actors (state forces, rebel groups, violent extremist organizations, identity and political militias, rioters, protesters, civilians, and external or other forces) based on their organizational structure, political or communal objectives, and relationships to communities. This classification enables the analysis of both formal and informal armed actors, from state militaries and rebel movements to ethnic militias and self-defence groups, offering a nuanced view of how political violence manifests across different territorial and social contexts.

For analytical purposes, the ACLED dataset groups events into distinct categories of violent interaction. In line with previous studies (OECD/SWAC, 2020; 2022; 2023; 2025), we use three of these categories that provide a spatially detailed and comparable basis for assessing patterns and intensities of armed conflict over time: battles, explosions and remote violence, and violence against civilians. Battles represent direct confrontations between armed actors and are the deadliest type of event, accounting for the majority of recorded fatalities. Explosions and remote violence encompass long-range or indiscriminate attacks such as bombings, airstrikes, or shelling, while violence against civilians captures attacks intentionally targeting unarmed populations by any organized armed group. Nonviolent actions such as protests or riots are also tracked by ACLED but are excluded from the main analysis of political violence in this paper, which focuses specifically on armed conflict dynamics.

Between 1 January 2012 and 30 June 2025, ACLED records 61,855 events that occurred with the 17 West Africa countries and that met the definitions described above. These events are responsible for an estimated 196,469 deaths. Our analysis focuses on the number of violent events, which is far less disputed than the number of deaths. For clarity, only the locations of events in the region during 2024-25 are shown in Map 1.

Map 1. Violent events in West Africa, 2024-mid 2025

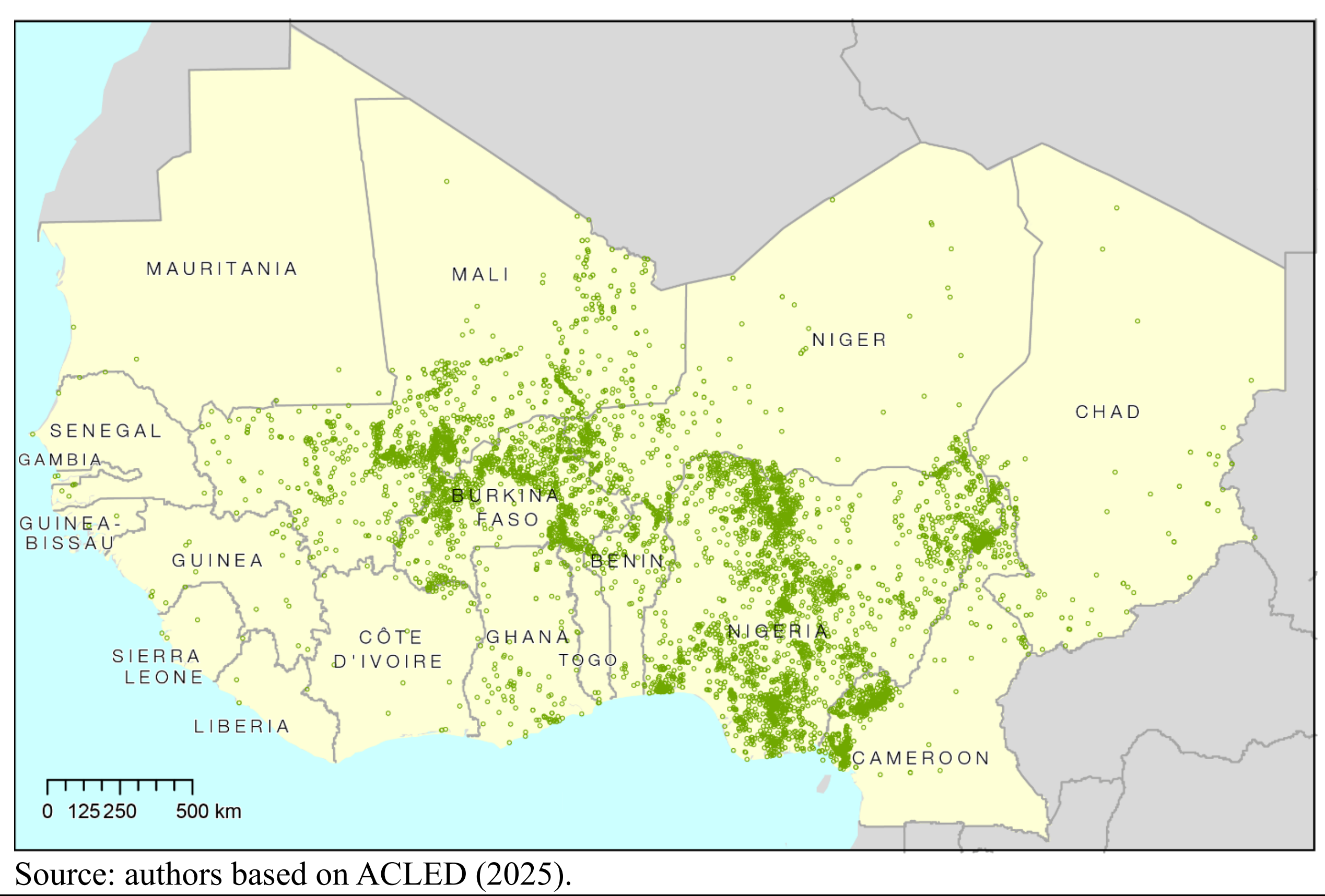

Source: authors based on ACLED (2025).

*Africapolis*

The SWAC Africapolis project is a continent-wide initiative to provide a geospatially-explicit database of urban agglomerations in Africa (OECD et al., 2025). It addresses the long-standing challenge of varied national definitions of “urban” areas, patchy demographic censuses, and an over-focus on large metropolitan centres. Africapolis adopts a uniform cross-national focus on urban agglomerations as those built-up areas with a permanent population of 10,000 or more and where buildings are less than 200 meters apart.

The dataset combines official population censuses with satellite and aerial imagery to map the extent of urban settlements across all of Africa. The spatial dataset consists of three primary vector layers: one each for the reference years of 2015, 2020, and 2025. Each layer contains urban agglomeration polygons that are linked with population estimates. The perimeters for 2015 and 2020 were calculated using satellite images from Google Earth, while the perimeter for 2025 is a projection based on previous years. For calculation purposes, the built-up area is rasterized, which can sometimes lead to an overestimation of built-up area, particularly for small agglomerations, and some small discrepancies between 2020 and 2025. Similarly, the population figures for 2015 and 2020 come from national census data, while the data for 2025 are projected from previous years. When census data for either 2015 or 2020 is missing, data are extrapolated from the closest available census year based on country-level growth rates.

Within the set of 17 West Africa countries covered by this paper, Africapolis records several thousand agglomerations each of the three reference years: 2,903 in 2015, 3,916 in 2020, and 4,277 for 2025. The increase in the number of agglomerations over time reflects the region's rapid population growth and urbanization. The region's 4,200+ agglomerations for 2025 are shown in Map 2.

Map 2. Urban agglomerations in West Africa, 2025

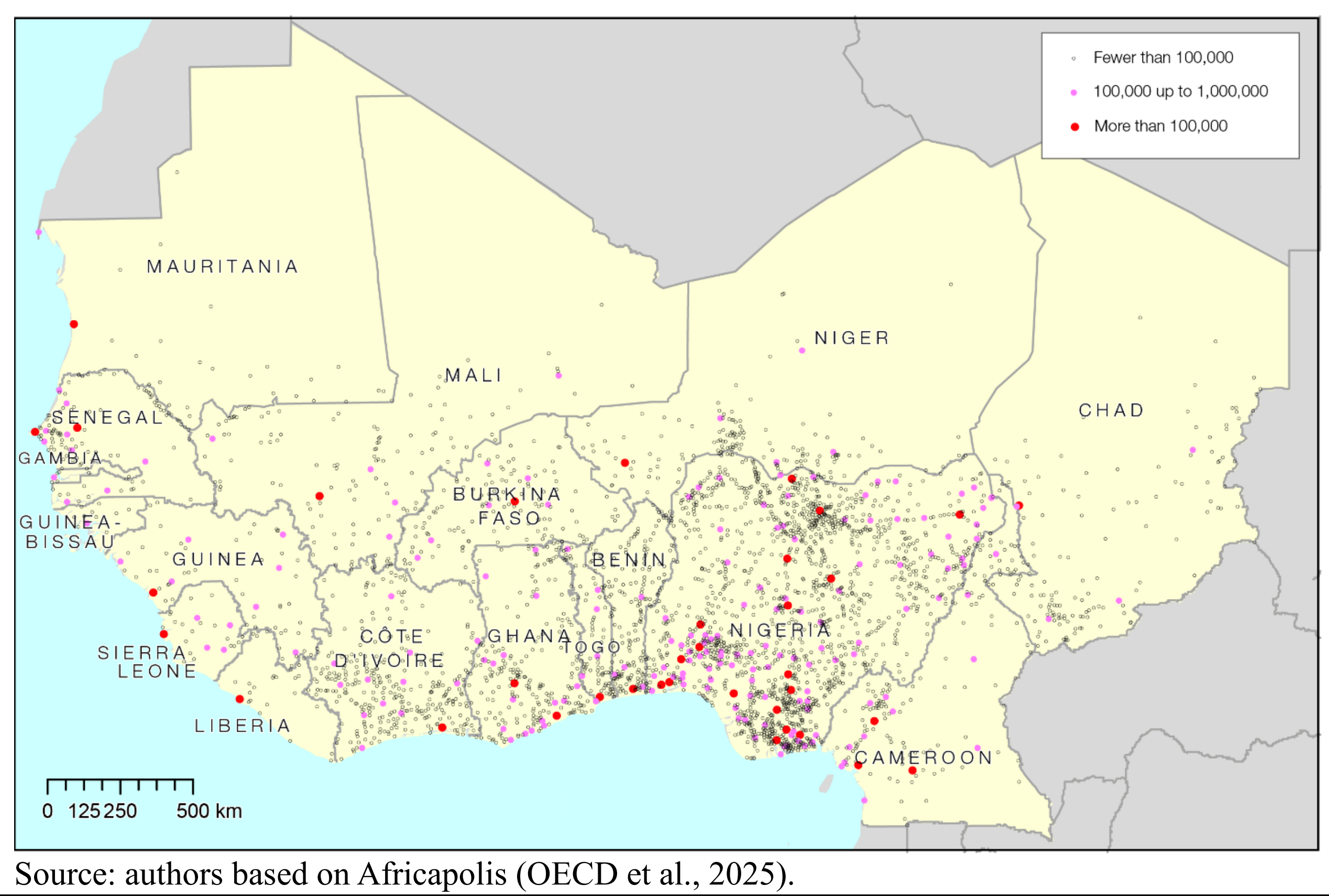

Source: authors based on Africapolis (OECD et al., 2025).

*Combining conflict and population data*

First, violent events and urban agglomerations were combined in a Geographic Information System (GIS) to identify the location of events relative to the nearest agglomeration. More specifically, the name of the nearest agglomeration was identified for each event and the distance in kilometres between the event and the nearest agglomeration was also calculated. Events that occurred within the area of an agglomeration were coded at a distance of 0. Because agglomerations are only available for three years (2015, 2020, and 2025), agglomeration names and event distances were calculated for three time periods based on the associations described in Table 1.

Table 1. Association between urban agglomerations with violent events

| Africapolis agglomeration year | ACLED event years |
|---|---|
| 2015 | 2012-2016 |
| 2020 | 2017-2021 |
| 2025 | 2022-2025* |

Sources: ACLED (2025) and Africapolis (OECD et al., 2025).
Note: *partial year data through 30 June.

Second, relating each event to the nearest agglomeration also allowed the calculation of a ratio measurement that compares the number of events that occur outside the area of an agglomeration to those that occur within it. Following Gizelis et al. (2021), outside events were included in peri-urban areas only up to a distance of 30 km (Figure 1).

Figure 1. Defining urban, peri-urban, and rural zones for the analysis

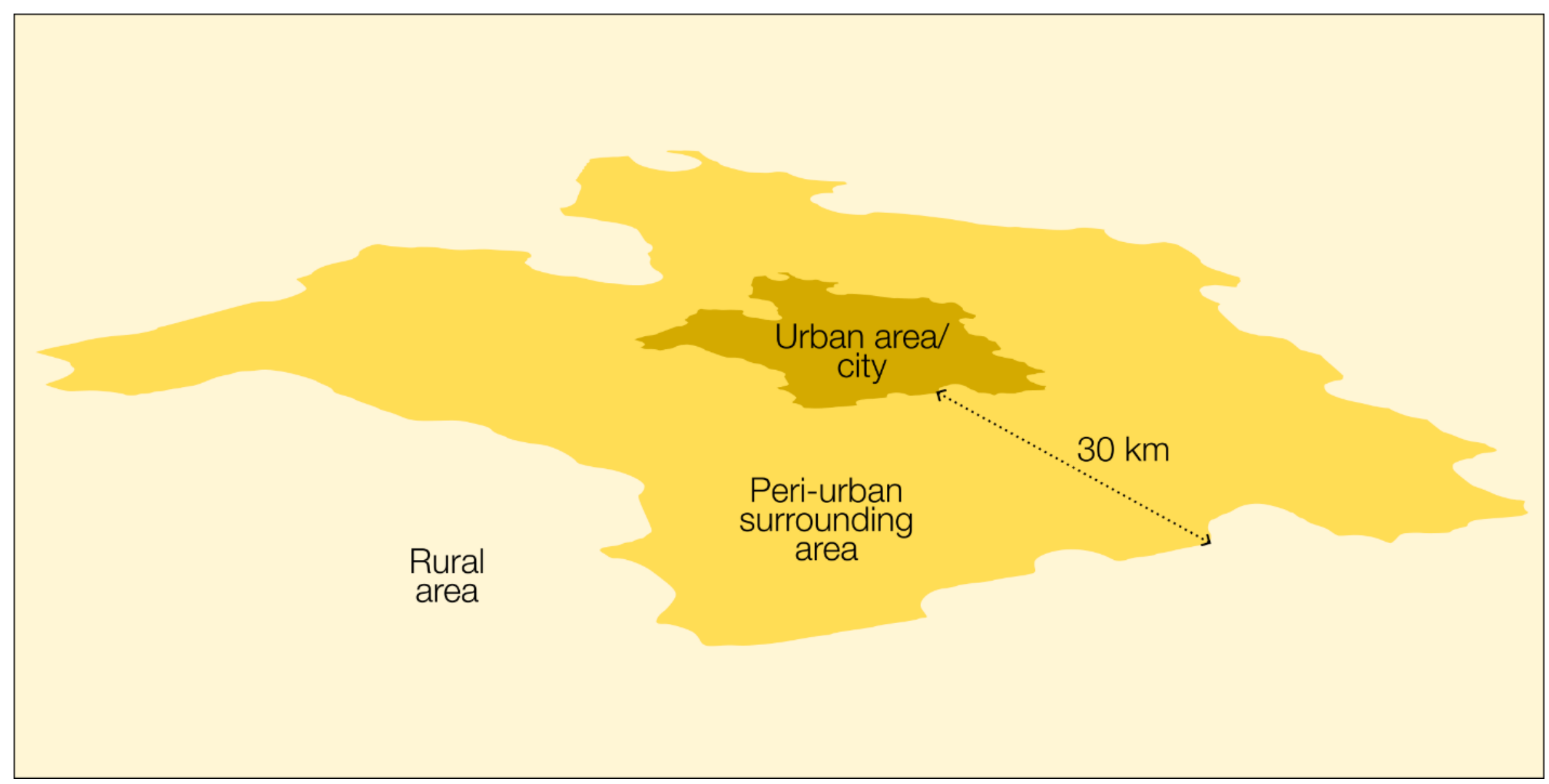

This resulting score for each agglomeration reflects the spatial patterning of events relative to agglomeration's area (Table 2). Based on these patterns, the analysis grouped each category into two main types: those agglomerations with more violence outside and those with more within.

Table 2. Spatial patterning of violent events relative to agglomeration

| **Outside: within ratio score** | **Interpretation** | **Classification for the analysis** |
|---|---|---|
| 0 | Some violence outside an agglomeration, none within | More violence outside |
| > 0 up to 1 | More violence outside than within | More violence outside |
| 1 | Same amount of violence outside as within | More violence within |
| > 1 | More violence within than outside | More violence within |
| Undefined | Some violence within, none beyond | More violence within |

Sources: authors based on ACLED (2025) and Africapolis (OECD et al., 2025).

Finally, the distance in kilometres from each agglomeration to the nearest international border was calculated to identify borderland agglomerations.

**More violent events near urban agglomerations**

The vast majority of violent events occur either within city boundaries or in their immediate peripheries. This proximity to urban agglomerations is clearly visible on Figure 2, which presents the spatial distribution of political violence in West Africa between 2012 and mid-2025, measured by distance from the nearest Africapolis urban agglomeration. Across West Africa, 41 percent of violence occurs within urban agglomerations, with an additional 44 percent occurring in areas extending up to 30 km beyond urban limits. This pattern reinforces previous findings that that while urban areas remain focal points of instability, West African violence is spatially diffuse (Walther et al., 2023), often spilling into peri-urban and rural belts. Nonetheless, the steep decline in event frequency beyond 30 km confirms that political violence is tightly coupled to urban systems.

Despite national differences in overall conflict intensity, a broadly consistent spatial pattern emerges: the majority of violent events occur inside urban boundaries, followed by a sharp decline in frequency within the first 30 km beyond city limits. This clear distance decay is represented on Figure 3, which illustrates how the share of political violence within and around urban agglomerations varies across individual West African countries between 2012 and 2025. Countries such as Burkina Faso, Nigeria, and Mali show somewhat flatter curves, indicating more substantial levels of peri-urban or rural violence relative to their urban cores, reflecting the diffusion of conflict along transport corridors and into hinterland settlements. In contrast, coastal states like Ghana, Guinea, and Senegal exhibit steeper declines, suggesting that violence remains overwhelmingly concentrated in urban centres. Overall, the figure demonstrates a shared urban anchor to political violence across West Africa, with meaningful variation in how far beyond city limits instability extends, reflecting differences in national territorial control and conflict geographies.

Figure 2. Violent events by distance from nearest urban agglomeration, in %, 2012-mid 2025

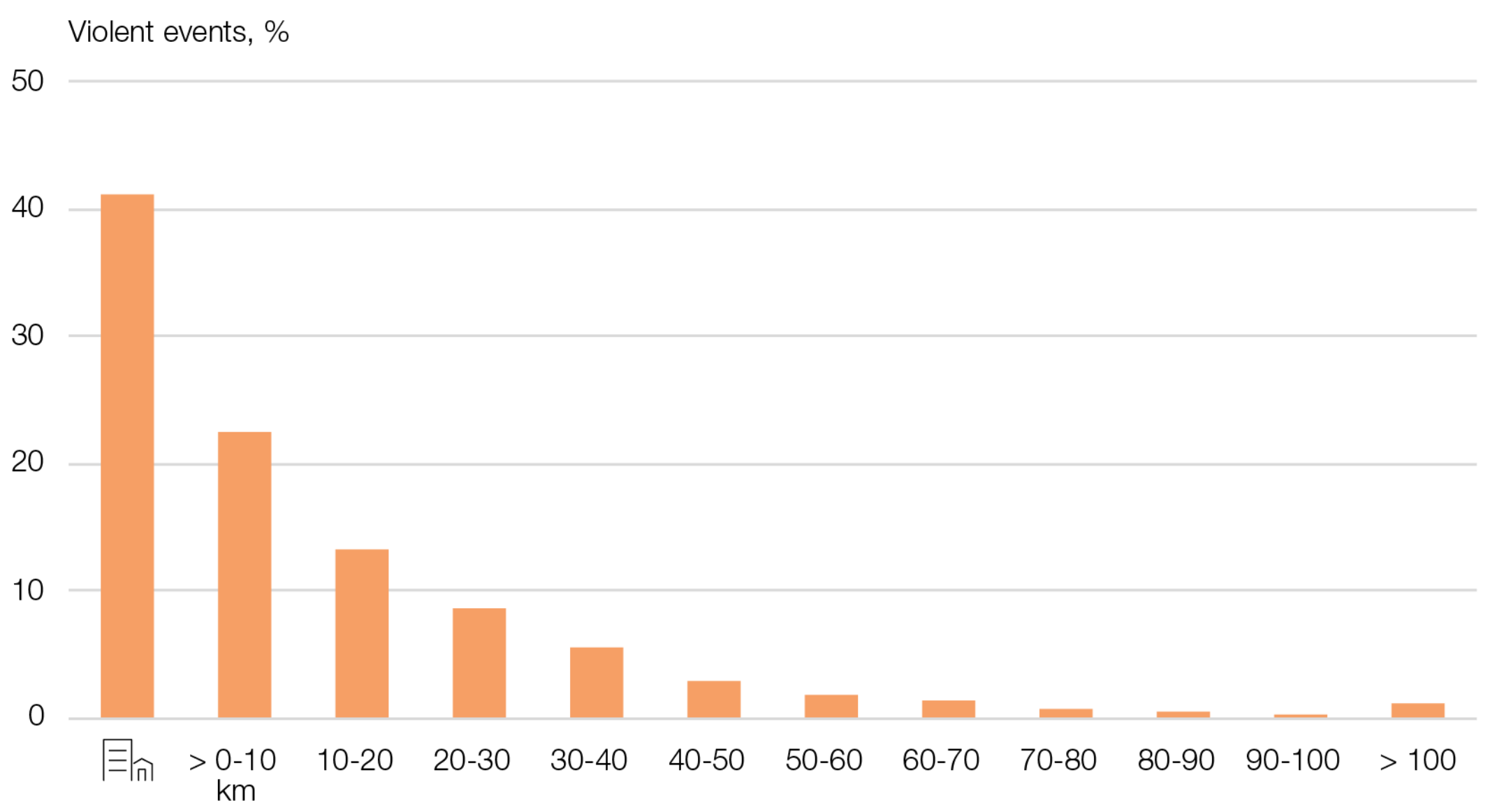

Sources: authors based on ACLED (2025) and Africapolis (OECD et al., 2025).

Figure 3. Violent events by distance from nearest urban agglomeration, by country, in %, 2012-mid 2025

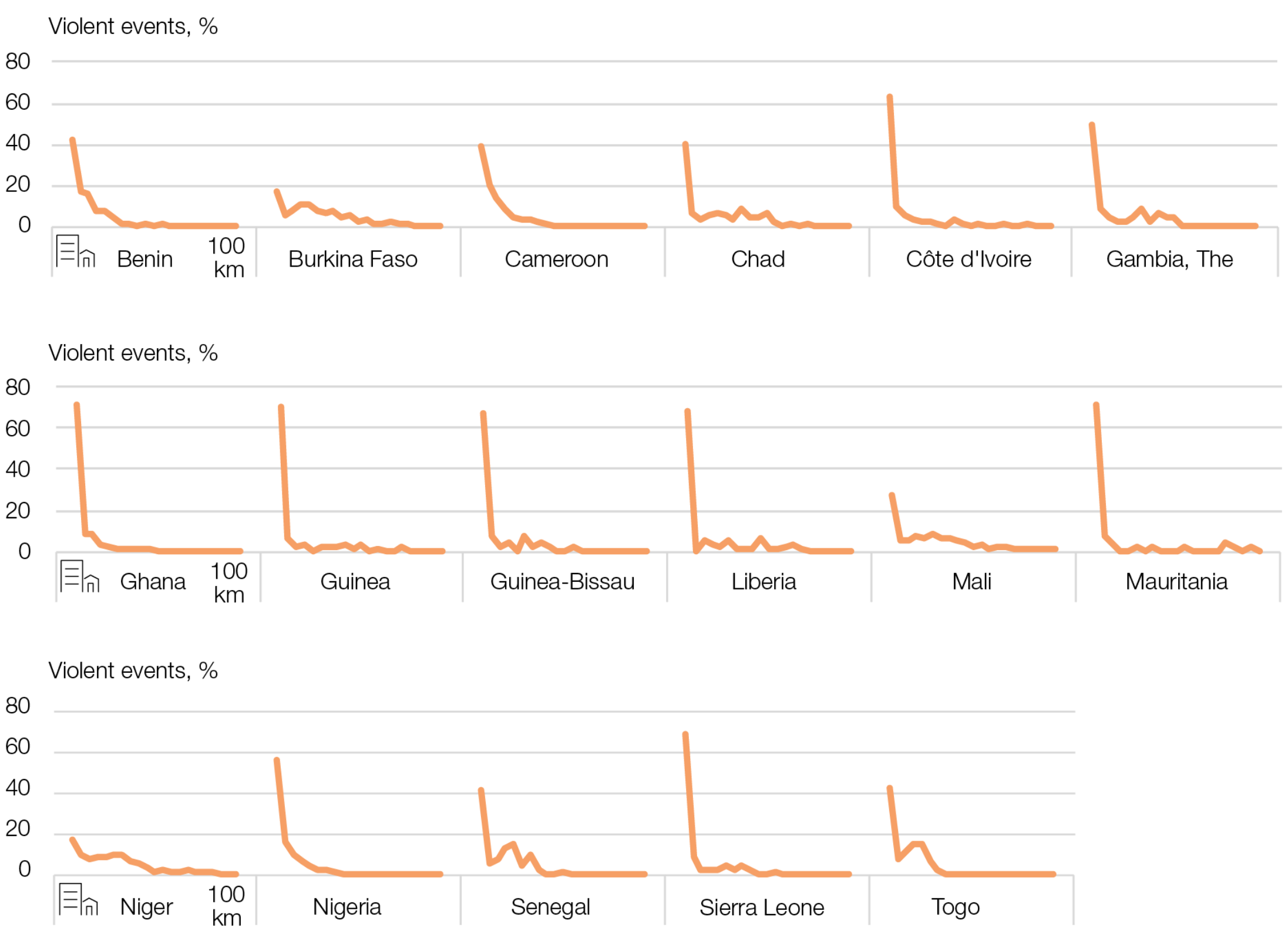

Sources: authors based on ACLED (2025) and Africapolis (OECD et al., 2025).

*Less urban violence over time*

Political violence is becoming less urban over time. This ruralization of violence is represented on Figure 4, which tracks how the spatial distribution of violent events in West Africa evolved between 2012 and mid-2025. At the start of the period, more than three-quarters of all incidents occurred within urban boundaries, but this share steadily declined to around 40–45 percent by 2024. Over the same period, the proportion of events taking place in peri-urban and rural zones (those located up to 30 km from urban agglomerations) expanded significantly. This trend suggests that conflicts that were once centred in major cities and their immediate surroundings have progressively dispersed across broader hinterlands. The widening distribution of violence implies both a growing reach of armed actors beyond urban strongholds and the erosion of states' ability to contain insecurity within urban cores. Nonetheless, even in 2024, urban and peri-urban zones together still account for most violent events, highlighting the persistent centrality of cities in the region's conflict dynamics.

Figure 4. Violent events by distance from nearest urban agglomerations in West Africa, in %, 2012-2024

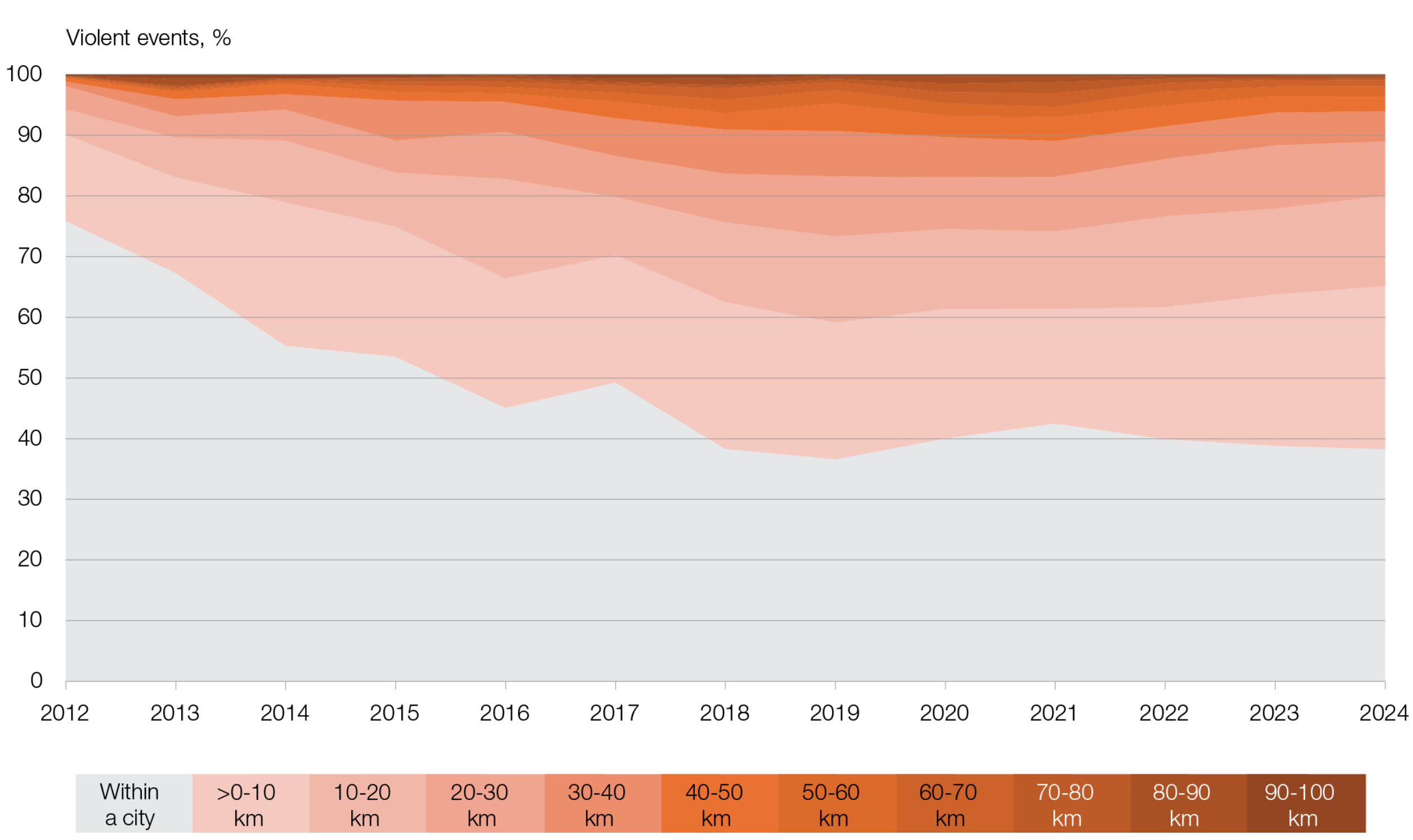

Sources: authors based on ACLED (2025) and Africapolis (OECD et al., 2025).

Our analysis also reveals that urban agglomerations with violence concentrated within their limits have the highest average population size. The fact that larger urban centres attract and sustain the majority of political violence is clearly visible on Figure 5, which compares the ratio of violent events occurring within urban agglomerations to those within a 30-km buffer surrounding them, classified by the average population of each agglomeration in 2025. Only a small share of urban agglomerations have either balanced levels of violence inside and outside or higher levels beyond the city edge, and these are typically smaller settlements. A modest number of agglomerations experience no recorded violence in either zone. Overall, the figure suggests that intra-urban is most pronounced in populous urban agglomerations where state presence, economic activity, and political competition are concentrated, while smaller towns and peri-urban zones remain comparatively less affected.

Figure 5. Ratio of violent events outside and within West African urban agglomerations, 2012-25

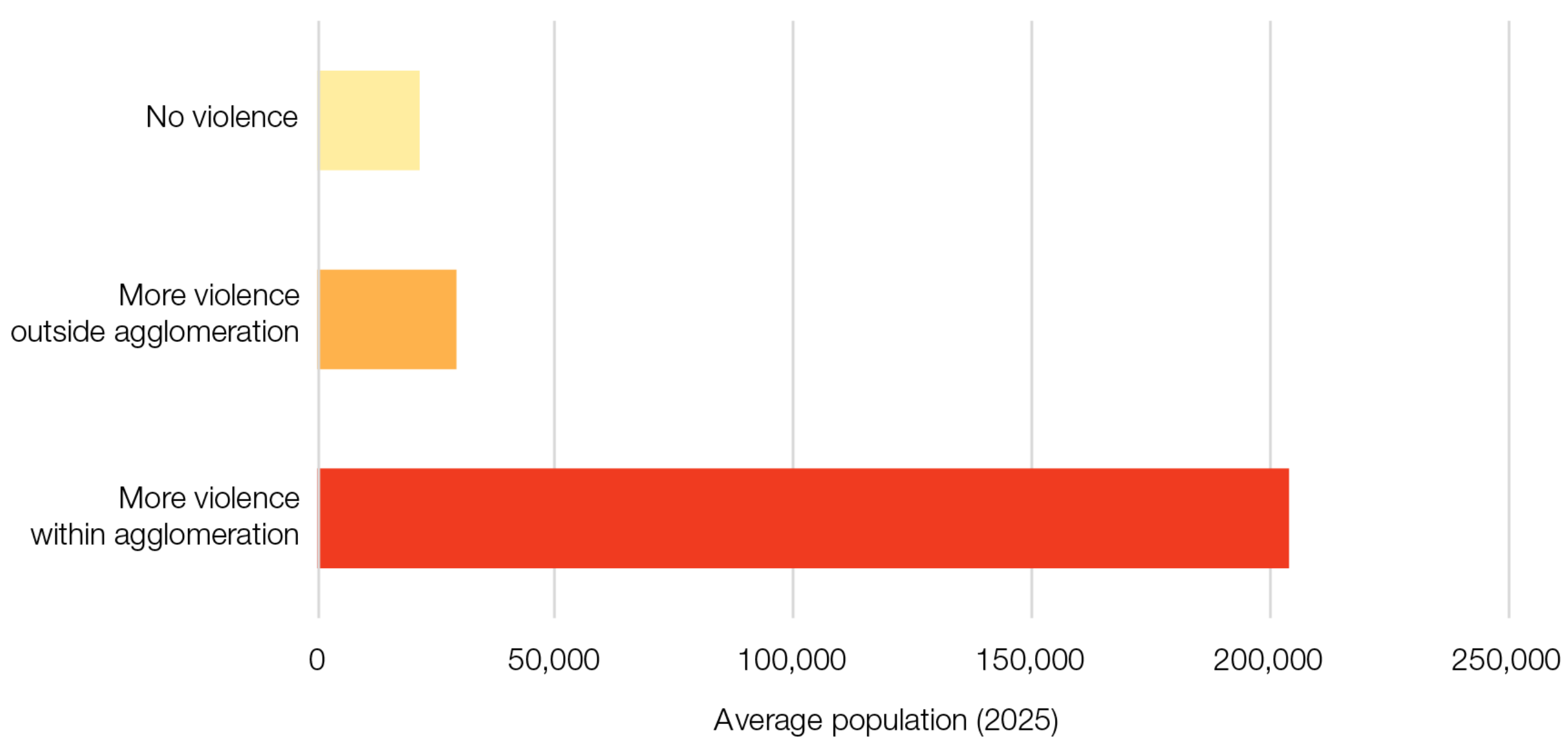

Sources: authors based on ACLED (2025) and Africapolis (OECD et al., 2025).
Note: we use a threshold of 30 km to distinguish between within and outside cities.

A clear spatial concentration of political violence is observed within a limited number of urban agglomerations in Burkina Faso, Cameroon, Mali, and Nigeria (see Table A1 in Appendix). Across the entire period, Cameroonian agglomerations dominate the rankings, with Bamenda recording the highest total number of violent events (1,534 incidents) when combining occurrences inside the city and within a 30-km buffer. Other Cameroonian agglomerations such as Kumbo, Mora, and Mayo-Moskota also register very high totals, indicating that urban conflict in Cameroon has been both intense and spatially extensive, spilling beyond city limits into adjacent rural zones. In Nigeria, Onitsha, Maiduguri, and Lagos stand out as major centres of violence within urban agglomerations, each recording several hundred to over a thousand events. These results reflect both the population size and the roles of cities as strategic or symbolic hubs in national conflict dynamics.

When considering violence beyond the city edge, the data show that smaller urban centres, including Mayo-Moskota, Mora (Cameroon) and Djibo (Burkina Faso), often experience proportionally higher levels of conflict in surrounding areas than within their built-up cores (see Table A2 in Appendix). This pattern suggests that many violent incidents occur along access roads, peri-urban settlements, and transitional zones between towns and the countryside. The inclusion of both within- and beyond-city counts underscores how political violence in West Africa is not confined to the largest metropolitan areas but is also pervasive in smaller cities where armed groups exploit limited state presence. Collectively, the results reveal that the geography of urban violence is highly uneven. It is concentrated in a handful of key national and regional centres while also diffusing outward along urban peripheries that serve as active frontiers of insecurity (Table 3).

Table 3. Urban agglomerations most affected by violence in West Africa, total, 2025

| *City* | *Country* | *Population, 2025* | *Urban population rank, 2025* | *Events within the city* | *Events outside the city (30 km)* | *Total* |
|---|---|---|---|---|---|---|
| Bamenda | Cameroon | 413,450 | 74 | 1,251 | 283 | 1,534 |
| Onitsha | Nigeria | 10,348,968 | 2 | 936 | 104 | 1,040 |
| Mayo-Moskota/ Ashigashiya | Cameroon | 12,007 | 2,795 | 90 | 682 | 772 |
| Mora | Cameroon | 51,902 | 545 | 228 | 496 | 724 |
| Maiduguri | Nigeria | 2,391,377 | 19 | 586 | 106 | 692 |
| Kumbo | Cameroon | 206,082 | 138 | 177 | 504 | 681 |
| Bula Chirabe/ Banki | Nigeria | 43,873 | 645 | 94 | 481 | 575 |
| Lagos | Nigeria | 16,108,180 | 1 | 493 | 20 | 513 |
| Djibo | Burkina Faso | 42,544 | 666 | 197 | 239 | 436 |
| Gao | Mali | 150,942 | 181 | 311 | 92 | 403 |
| Kumba | Cameroon | 217,494 | 130 | 116 | 281 | 397 |
| Buea | Cameroon | 194,474 | 147 | 312 | 77 | 389 |
| Ekombe Bonji | Cameroon | 17,820 | 1,801 | 82 | 284 | 366 |
| Ekondo-Titi | Cameroon | 17,347 | 1,865 | 89 | 260 | 349 |
| Bandiagara | Mali | 27,421 | 1,087 | 55 | 290 | 345 |
| Muyuka | Cameroon | 42,283 | 672 | 122 | 220 | 342 |
| Sofara | Mali | 15,888 | 2,066 | 21 | 297 | 318 |
| Port Harcourt | Nigeria | 3,773,353 | 12 | 296 | 22 | 318 |
| Batibo | Cameroon | 21,424 | 1,439 | 71 | 246 | 317 |
| Bama | Nigeria | 220,600 | 128 | 247 | 60 | 307 |

Sources: authors based on conflict data from ACLED (2025) and population data from Africapolis (OECD et al., 2025). The urban population rank is calculated for West Africa only.

*More violence in border regions*

Overall, urban agglomerations located within 10 to 30 km of a border are the most likely to experience more violence in their surrounding areas than inside their own boundaries. These findings suggest that border proximity involves the outward diffusion of violence into rural or cross-border zones. Figure 6 explores how this relationship between urban and peri-urban violence varies according to a city's population size as well as its distance from the nearest international border.

Smaller population borderland agglomerations experience much more outside violence up to 70 km away from borders. Beyond that distance, there is a more balanced or even reversed relationship, where violence within the city becomes relatively more common. However, for the largest agglomerations, violence is overwhelming internal no matter if it is in a border region or not.

Figure 6. Violent events in urban agglomerations by distance to nearest border, in %, 2025

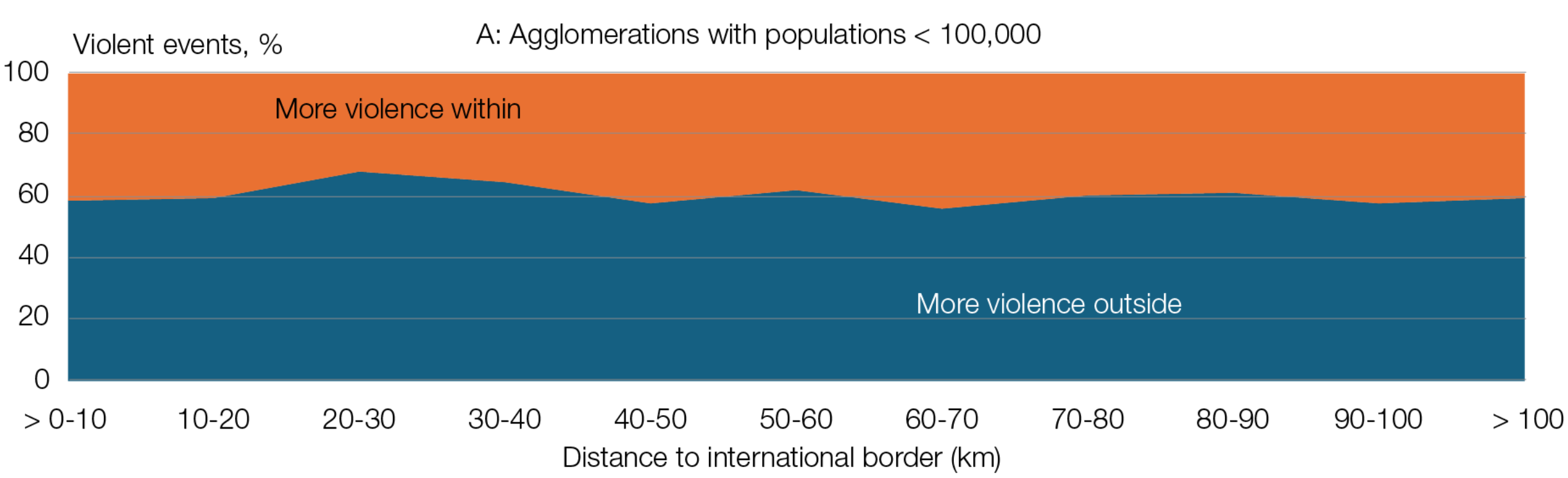

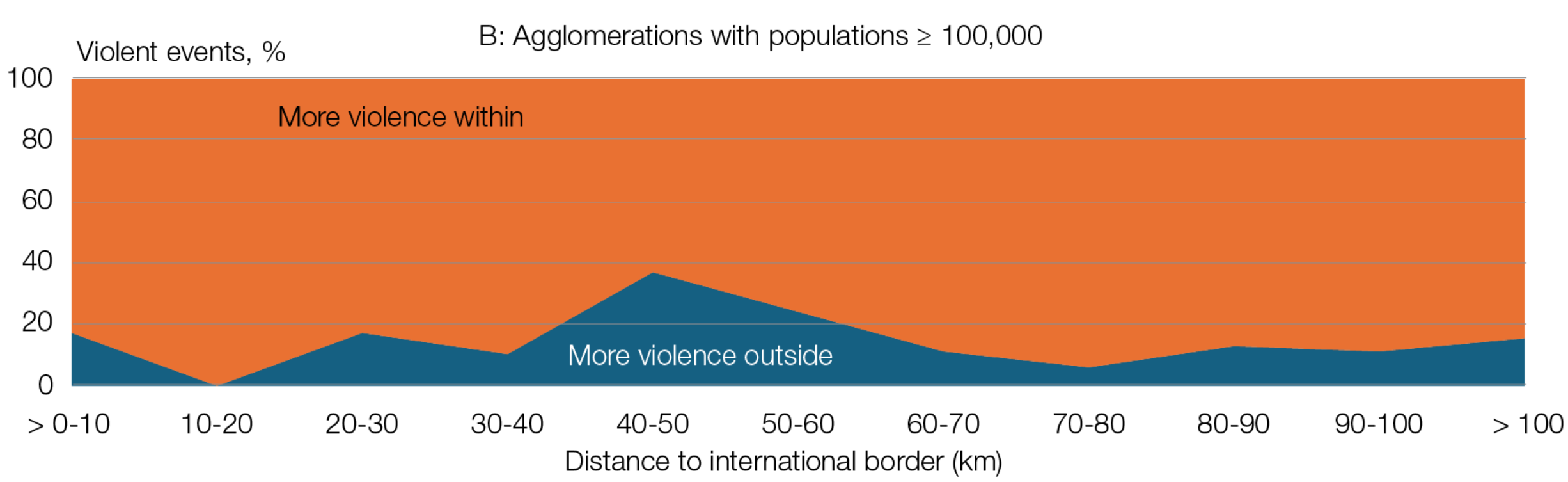

Sources: authors based on ACLED (2025) and Africapolis (OECD et al., 2025).

By considering population, the spatial logic of borderland violence becomes clearer. Smaller cities, which are more frequently situated near borders (OECD, 2019), overwhelmingly experience higher levels of violence outside their boundaries. This likely reflects weaker state presence, porous frontiers, and the mobility of armed actors who operate in borderlands. Larger urban agglomerations, on the other hand, tend to have violence concentrated within their own limits, reflecting both their demographic weight and their central political and economic roles.

These findings indicate that the geography of urban violence in West Africa depends jointly on proximity to borders and urban scale. Smaller borderland towns act as nodes of diffuse and mobile violence, while larger interior urban agglomerations remain the primary sites of politically driven or state-targeted conflict. For example, attacks on security forces, violence directed at public officials or journalists, or bombings of crowded public places like markets should be expected to have an urban focus as that is where most such targets are located.

**Discussion and policy implications**

This analysis has sought to gain a better understanding of the geographical distribution of political violence in West Africa by focusing on violent events observed in urban and rural areas. Each of these two environments offers very different resources to armed groups fighting against central governments. Cities are home to the symbols and institutions that insurgents hope to conquer in order to overthrow the government. Urban areas also make it possible to mobilize a more educated and politicized population than in rural areas.

It is therefore not surprising that, on a regional scale, cities and their immediate suburbs account for a significant proportion of the violence observed in the region since 2012. The article also confirms that violence decreases very rapidly as one moves away from urban areas, confirming a trend already observed in the early 2020s (OECD/SWAC, 2023). This marked distance decay observed at the regional level is also present in most West African countries. Burkina Faso, Mali, and Nigeria are exceptions to the rule, because a significant portion of their rural areas are currently in the hands of jihadist insurgents.

These Sahelian countries illustrate the fact that rural areas offer an unparalleled refuge for armed groups, where they can mobilize numerous agricultural and pastoral resources. Far from the central government, it is in the countryside that insurgents can move freely and force rural populations to adopt their modes of government inspired by a literal interpretation of religious texts. Over time, these movements have tended to become more rural. In the Central Sahel and around Lake Chad, most of the violence initially affected urban agglomerations, then the countryside. As a result, the proportion of violent events occurring in urban areas declined from three-quarters to less than half between 2012 and 2024 at the regional level.

Our analysis also shows that urban agglomerations are affected by violence in different ways depending on their size and geographical location within the country. In large urban agglomerations, a greater proportion of violence is observed in city centres than in their suburbs. The opposite is true for small towns, where peripheral areas are more violent than built-up cores. Peripheral violence is particularly intense around small border towns. These results confirm previous studies showing that small towns and border regions are on average more affected by armed conflict than other regions (OECD/SWAC, 2022; 2023).

These results lead to several policy implications.

First, the current ruralisation of armed conflicts observed in the Sahel challenges the *Maoist model of insurgency*, which assumes that insurgents must necessarily win over the countryside before conquering the cities. On the contrary, this paper suggests that armed groups use the comparative advantages of cities and the countryside according to the offensives carried out against them by states and to their own strategy of territorial conquest. In other words, the movements of armed groups between urban and rural areas should not only be interpreted as a consequence of counterinsurgency initiatives but also as an opportunist or strategic move.

Non-state actors have their own agency when it comes to fighting in urban or rural areas. Thus far Jihadist groups have imposed numerous blockades on key cities, such as Kayes, Djibo, Farabougou, Timbuktu, Kayes, Nioro du Sahel, and, more recently, Bamako. It is unclear whether this strategy should lead to a third stage of the insurgency, during which Jihadist groups try to occupy cities militarily. Thus far, very little fighting has taken place within major urban areas in the Sahel, as government forces, rebels, and jihadists preferred to withdraw from cities in the face of their opponents' advance.

Second, the ruralisation of West African conflicts also suggests that particular attention should be paid to protecting civilians, who are the main victims of violence in rural areas. This protection involves the large-scale deployment of small mobile units in rural areas, a strategy that contrasts with the current trend of grouping government forces in fortified camps between which armoured convoys circulate. However, improving security in West Africa will require investing not only in troops, but in transport infrastructure as the lack of accessible transport remains one of the biggest obstacles to both security and development. Expanding and maintaining road networks could reduce isolation, facilitate trade, and allow faster security and humanitarian responses. Sixty years after independence, vast portions of the region still lack paved roads connecting them to national capitals. This absence leaves populations feeling abandoned and provides fertile ground for rebellion. As long as isolation persists, insurgents will continue to exploit this geography of neglect.

A third implication, closely linked to the first, is to devote more attention and investment to cities in border regions. The past decade has clearly demonstrated the strategic importance of national peripheries for national cohesion and the usefulness of improving infrastructure that connect them to national centres. Developing peripheral areas remains the surest way to secure the region, particularly through border towns, which play a key role in the movement of people and goods between West African countries. Rather than relying only on approaches that prioritize securitizing border areas, more investment should be dedicated to supporting production facilities and border infrastructure, such as markets, juxtaposed checkpoints, major roads, and educational and medical services.

## Appendix

Table A1. Urban agglomerations most affected by within violence in West Africa, 2025

| *City* | *Country* | *Population, 2025* | *Urban population rank, 2025* | *Events within the city* | *Events beyond the city (up to 30 km)* | *Total* |
|---|---|---|---|---|---|---|
| Bamenda | Cameroon | 413,450 | 74 | 1,251 | 283 | 1,534 |
| Onitsha | Nigeria | 10,348,968 | 2 | 936 | 104 | 1,040 |
| Maiduguri | Nigeria | 2,391,377 | 19 | 586 | 106 | 692 |
| Lagos | Nigeria | 16,108,180 | 1 | 493 | 20 | 513 |
| Buea | Cameroon | 194,474 | 147 | 312 | 77 | 389 |
| Gao | Mali | 150,942 | 181 | 311 | 92 | 403 |
| Port Harcourt | Nigeria | 3,773,353 | 12 | 296 | 22 | 318 |
| Kidal | Mali | 52,418 | 535 | 260 | 18 | 278 |
| Bama | Nigeria | 220,600 | 128 | 247 | 60 | 307 |
| Benin City | Nigeria | 1,843,500 | 26 | 243 | 24 | 267 |
| Mora | Cameroon | 51,902 | 545 | 228 | 496 | 724 |
| Damboa | Nigeria | 220,996 | 127 | 222 | 62 | 284 |
| Timbuktu | Mali | 89,664 | 301 | 206 | 23 | 229 |
| Kaduna | Nigeria | 2,378,538 | 20 | 202 | 46 | 248 |
| Djibo | Burkina Faso | 42,544 | 666 | 197 | 239 | 436 |
| Menaka | Mali | 14,878 | 2,235 | 191 | 56 | 247 |
| Kumbo | Cameroon | 206,082 | 138 | 177 | 504 | 681 |
| Jos | Nigeria | 1,269,635 | 34 | 175 | 53 | 228 |
| Monguno | Nigeria | 410,839 | 75 | 172 | 46 | 218 |
| Gwoza | Nigeria | 122,002 | 221 | 167 | 79 | 246 |

Sources: authors based on conflict data from ACLED (2025) and population data from Africapolis (2025). The urban population rank is calculated for West Africa only.

Table A2. Urban agglomerations most affected by peripheral violence in West Africa, 2025

| *City* | *Country* | *Population, 2025* | *Urban population rank, 2025* | *Events within the city* | *Events beyond the city (up to 30 km)* | *Total* |
|---|---|---|---|---|---|---|
| Mayo-Moskota/ Ashigashiya | Cameroon | 12,007 | 2,795 | 90 | 682 | 772 |
| Kumbo | Cameroon | 206,082 | 138 | 177 | 504 | 681 |
| Mora | Cameroon | 51,902 | 545 | 228 | 496 | 724 |
| Bula Chirabe/Banki | Nigeria | 43,873 | 645 | 94 | 481 | 575 |
| Sofara | Mali | 15,888 | 2,066 | 21 | 297 | 318 |
| Dikwa | Nigeria | 192,993 | 152 | 0 | 297 | 297 |
| Bandiagara | Mali | 27,421 | 1,087 | 55 | 290 | 345 |
| Ekombe Bonji | Cameroon | 17,820 | 1,801 | 82 | 284 | 366 |
| Bamenda | Cameroon | 413,450 | 74 | 1,251 | 283 | 1,534 |
| Kumba | Cameroon | 217,494 | 130 | 116 | 281 | 397 |
| Koza | Cameroon | 25,964 | 1,160 | 29 | 275 | 304 |
| Ekondo-Titi | Cameroon | 17,347 | 1,865 | 89 | 260 | 349 |
| Batibo | Cameroon | 21,424 | 1,439 | 71 | 246 | 317 |
| Blangoua | Cameroon | 13,512 | 2,463 | 0 | 244 | 244 |
| Falagountou | Burkina Faso | 11,032 | 3,035 | 0 | 242 | 242 |
| Djibo | Burkina Faso | 42,544 | 666 | 197 | 239 | 436 |
| Damasak | Nigeria | 91,030 | 293 | 0 | 228 | 228 |
| Muyuka | Cameroon | 42,283 | 672 | 122 | 220 | 342 |
| Tori | Mali | 15,205 | 2,184 | 3 | 203 | 206 |
| Diffa | Niger | 63,014 | 443 | 87 | 188 | 275 |

Sources: authors based on conflict data from ACLED (2025) and population data from Africapolis (2025). The urban population rank is calculated for West Africa only.

## References

ACLED (2025). Armed Conflict Location & Event Data Project, https://acleddata.com

Beall, J., Goodfellow, T., & Rodgers, D. (2013). Cities and conflict in fragile states in the developing world. *Urban Studies*, *50*(15), 3065-3083.

Buhaug, H., & Urdal, H. (2013). An urbanization bomb? Population growth and social disorder in cities. *Global Environmental Change*, *23*(1), 1-10.

Büscher, K. (2020). African cities and violent conflict: the urban dimension of conflict and post conflict dynamics in Central and Eastern Africa. *Journal of Eastern African Studies*, *12*(2), 193–210.

Dorward, N., & Fox, S. (2022). Population pressure, political institutions, and protests: A multilevel analysis of protest events in African cities. *Political Geography*, *99*, 102762.

Dorward, N. (2024). The urbanization of conflict? Patterns of armed conflict and protest in Africa. *African Affairs*, *123*(493), 468-501.

Elfversson, E. (2025). Contentious cities? Urban growth and electoral violence in Africa. *World Development*, *193*, 107066.

Elfversson, E., & Höglund, K. (2021). Are armed conflicts becoming more urban?. *Cities*, *119*, 103356.

Elfversson, E., Höglund, K., Sellström, A. M., & Pellerin, C. (2023). Contesting the growing city? Forms of urban growth and consequences for communal violence. *Political Geography*, *100*, 102810.

Fox, S., & Bell, A. (2016). Urban geography and protest mobilization in Africa. *Political Geography*, *53*, 54-64.

Gizelis, T. I., Pickering, S., & Urdal, H. (2021). Conflict on the urban fringe: Urbanization, environmental stress, and urban unrest in Africa. *Political Geography*, *86*, 102357.

Golooba-Mutebi, F., & Sjögren, A. (2017). From rural rebellions to urban riots: political competition and changing patterns of violent political revolt in Uganda. *Commonwealth & Comparative Politics*, *55*(1), 22-40.

Goodfellow, T., & Jackman, D. (2023). *Controlling the capital: Political dominance in the urbanizing world*. Oxford University Press.

Hendrix, C. S. (2011). Head for the hills? Rough terrain, state capacity, and civil war onset. *Civil Wars*, *13*(4), 345-370.

Kaldor, M., & Sassen, S. (Eds.). (2020). *Cities at war: Global insecurity and urban resistance*. Columbia University Press.

Kniknie, S., & Büscher, K. (2023). Rebellious riots: entangled geographies of contention in Africa. *Rebellious Riots: Entangled Geographies of Contention in Africa*, 1-22.

Mkandawire, T. (2002). The terrible toll of post-colonial 'rebel movements' in Africa: towards an explanation of the violence against the peasantry. *The Journal of Modern African Studies*, *40*(2), 181-215.

Nedal, D., Stewart, M., & Weintraub, M. (2020). Urban concentration and civil war. *Journal of Conflict Resolution*, *64*(6), 1146-1171.

OECD/SWAC (2025), *Roads and Conflicts in North and West Africa*, West African Studies, OECD Publishing, Paris, https://doi.org/10.1787/77474489-en.

OECD/SWAC (2023), *Urbanisation and Conflicts in North and West Africa*, West African Studies, OECD Publishing, Paris, https://doi.org/10.1787/3fc68183-en.

OECD/SWAC (2022), *Borders and Conflicts in North and West Africa*, West African Studies, OECD Publishing, Paris, https://doi.org/10.1787/6da6d21e-en.
OECD/SWAC (2020), *The Geography of Conflict in North and West Africa*, West African Studies, OECD Publishing, Paris, https://doi.org/10.1787/02181039-en.
OECD et al. (2025), *Africa's Urbanisation Dynamics 2025: Planning for Urban Expansion*, West African Studies, OECD Publishing, Paris, https://doi.org/10.1787/2a47845c-en.
OECD (2019), "Population and Morphology of Border Cities", *West African Papers*, No. 21, OECD Publishing, Paris, https://doi.org/10.1787/80dfd9d8-en.
Østby, G. (2016). Rural–urban migration, inequality and urban social disorder: Evidence from African and Asian cities. *Conflict Management and Peace Science*, *33*(5), 491-515.
Peterson, T. G. (2024). *Revolutionary warfare: How the Algerian war made modern counterinsurgency*. Cornell University Press.
Radil, S. M., Walther, O., Dorward, N., & Pflaum, M. (2023). Urban-rural geographies of political violence in North and West Africa. *African Security*, *16*(2-3), 199-222.
Radil, S. M., Irmischer, I., & Walther, O. J. (2022). Contextualizing the relationship between borderlands and political violence: A dynamic space-time analysis in North and West Africa. *Journal of Borderlands Studies*, *37*(2), 253-271.
Retaillé, D., & Walther, O. (2013). Conceptualizing the mobility of space through the Malian conflict. *Annales de Géographie*, 694(6), 595-618.
Thurston, A. (2020). *Jihadists of North Africa and the Sahel: Local politics and rebel groups*. Cambridge University Press.
Thurston, A. (2018). *Boko Haram: the history of an African jihadist movement*. Princeton University Press.
Walther, O. J., Radil, S. M., Russell, D. G., & Trémolières, M. (2023). Introducing the spatial conflict dynamics indicator of political violence. *Terrorism and political violence*, *35*(3), 533-552.
Walther, O. J., Radil, S. M., & Russell, D. G. (2025). The Spatial Conflict Life Cycle in Africa. *Annals of the American Association of Geographers*, *115*(2), 456-477.